\documentclass[5p]{elsarticle}

\usepackage{color}
\usepackage{hyperref}

\journal{Acta Materialia}

\biboptions{sort&compress}
\newcommand{\MPam}{MPa~m$^{1/2}$}

\begin{document}

\begin{frontmatter}

\title{On the influence of Al-concentration on the fracture toughness of NiAl: 
microcantilever fracture tests and atomistic simulations}



\author[ww1]{Ralf Webler}
\author[ww1]{Polina N. Baranova}
\author[ww1]{Shivraj Karewar}
\author[ww1]{Steffen Neumeier}
\author[ww1]{Johannes J. M\"oller}
\author[MPI]{Hauke Springer}
\author[ww1]{Mathias G\"oken}
\author[ww1]{Erik Bitzek\corref{mycorrespondingauthor}}
\cortext[mycorrespondingauthor]{Corresponding author}

\ead{erik.bitzek@fau.de}

\address[ww1]{Friedrich-Alexander-Universit\"at Erlangen-N\"urnberg, Department for Materials Science and Engineering, Institute I, Martenstr. 5, 91058 Erlangen, Germany}
\address[MPI]{Max-Planck-Institut f\"ur Eisenforschung GmbH, Department Microstructure Physics and Alloy Design, Max-Planck-Stra{\ss}e 1, 40237 D\"usseldorf, Germany}

\begin{abstract}
The mechanical properties of the stoichiometric B2 $\beta$-phase of NiAl are well established, however the effect of off-stoichiometric composition on the fracture toughness has not yet been systematically studied over the entire composition range of 40-50\% Al. 
Here we use microbending tests on notched cantilever beams FIB-milled from NiAl single crystals with an aluminized as well as an  oxidation-induced composition gradient to determine the influence of the Al concentration on the mechanical properties. 
The fracture toughness is maximal for the stoichiometric composition.  It decreases with increasing Ni-content in the Ni-rich composition range,  where  plastic deformation 
is observed to accompany the fracture process.
In contrast, no plasticity is observed in Al-rich NiAl, which
shows a nearly concentration-independent, low fracture toughness.
The theoretical fracture toughness according to Griffith, however, shows only a very weak composition dependence  in both, the Ni- and Al-rich composition range.   
The differences in fracture toughness could furthermore not 
be explained solely based on the different hardening contributions of Ni-antisites in the Ni-rich and structural vacancies in the Al-rich crystals. 
Atomistic fracture simulations show that crack propagation in  NiAl takes place by the nucleation and migration of kinks on the crack front.  
The low fracture toughness of Al-rich NiAl can thus be understood by the dual effect of structural vacancies as strong obstacles to dislocation motion and as source of crack front kinks.
\end{abstract}

\begin{keyword}

NiAl \sep fracture toughness \sep micro cantilever bending \sep atomistic simulation
%\texttt{elsarticle.cls}\sep \LaTeX\sep Elsevier \sep template
%\MSC[2010] 00-01\sep  99-00
\end{keyword}

\end{frontmatter}

%\listoftodos

%\linenumbers

\section{Introduction}
The intermetallic phase NiAl is regarded as promising candidate material for high temperature applications because of its low density, high thermal conductivity and excellent oxidation resistance \cite{Darolia1991,Miracle1993}. 
However, the low fracture toughness and poor ductility at room temperature as well as the comparatively low creep strength have been major obstacles towards its application as high temperature structural material \cite{Darolia2000}.  
NiAl, which crystalizes in the B2 (CsCl) structure, is however frequently used as a bond coat material for oxidation protection of turbine blades and vanes in aircraft engines \cite{Goward1998}. 
These bond coats have a characteristic composition gradient throughout their thickness, which is believed to significantly affect their mechanical behavior and failure resistance under load \cite{AranaAntelo1998b}. 
Understanding the composition dependent mechanical properties of NiAl is therefore of prime importance for improving the reliability of bond coats on turbine blades and vanes. 

The study of the composition dependence of fracture in intermetallic alloys is, however, also interesting from a fundamental point of view. 
The fracture toughness of semi-brittle materials like B2 NiAl is determined by the competition between bond breaking and dislocation processes at the crack tip \cite{Gumbsch2001}. 
While the influence of Al concentration on the yield stress and the dislocation mobility in NiAl has been extensively studied \cite{Baker1995,Pike1997,Pike2002}, including by atomistic simulations \cite{Schroll1997,Gumbsch1999,Schroll1998a,Schroll1998e,Medvedeva1998}, no such detailed studies exist  regarding  the bond breaking processes leading to brittle cleavage.

Since typical bond coats have thicknesses of just up to 100 $\mu$m, testing their mechanical properties with standard methods is challenging. 
In-situ microscale mechanical testing of cantilevers milled by a focused ion beam (FIB) as introduced by Di Maio and Halford \cite{DiMaio2005b,Hal05} can address these challenges and has been successfully applied to  determine the properties of materials like Si \cite{DiMaio2005b}, TiAl \cite{Hal05,Iqbal12}, W \cite{Wurster2012}, Cu \cite{Demir2010c}, diamond-like-carbon (DLC) \cite{Schaufler2012} and recently also of NiAl bond coats \cite{JayaB2012,Web2012,Jaya2014}. 
The study by Jaya et al. \cite{JayaB2012} was however conducted on polycrystalline parts of the bond coat so that grain boundary (GB) fracture dominated. 
The cantilevers in \cite{Web2012} were milled in individual grains, thus in principle allowing to compare their results to macroscopic fracture toughness on single crystalline NiAl, see e.g., \cite{Chang1992}. However, they only studied stoichiometric and Ni$_{60}$Al$_{40}$ coatings, which contained additional elements like Cr, Co and Pt. 

A different approach to study the influence of Al concentration on the fracture behavior of NiAl thin films was used by Wellner et al. \cite{Wellner2004}. 
They used thermal cycling to induce strain in polycrystalline NiAl films of different homogeneous compositions on a Si substrate.  Stoichiometric and Ni-rich NiAl films showed no cracks in their study, whereas Al-rich films  exhibited pronounced cracking. 
However, the fracture toughness could only be determined for Al-rich films, where GB fracture dominated.  
A systematic study on the influence of off-stoichiometric composition on the fracture toughness of pure, single-crystalline NiAl is currently lacking. 

By performing microcantilever tests with different cantilever sizes and geometries on stoichiometric NiAl, Ast et al. could establish that the fracture toughness of NiAl at room temperature does not depend on the specimen size for characteristic lengths down to the micro scale \cite{Ast14}. 
In-situ microscale testing on compositionally graded single-crystalline NiAl thus present a unique model system to study the effect of off-stoichiometric composition on the fracture behavior of the intermetallic B2 Ni$_{0.5-x}$Al$_{0.5+x}$ phase.

A complementary approach to study the influence of Al concentration on the fracture toughness of NiAl is the use of atomistic simulations. 
Atomistic simulations have provided considerable insight in the atomic scale details of brittle fracture and crack tip plasticity in general \cite{Bitzek2015a}. 
Previous molecular static (MS) simulations of cracks in NiAl \cite{Ludwig1998, Farkas2000} have confirmed the experimental finding that the \{110\} planes are the primary and most probably the only natural cleavage planes of NiAl \cite{Chang1992}. 
Dislocation emission was only observed for mixed mode loading or cracks on other planes. 
All emitted dislocations had $\langle 100 \rangle$ type Burgers vectors and were gliding on \{110\} planes \cite{Ludwig1998, Farkas2000}. 
Molecular dynamics (MD) simulations at 5\,K by Guo et al. \cite{Guo2007} suggest that at low temperatures and high strains the stress-induced formation of martensite might contribute to the dissipation of strain energy ahead of the crack front. 
However, unphysical phase transitions at crack tips are a notorious problem for potentials of the embedded atom method (EAM) type \cite{Moe14msmse,Moller2018}. 
All simulations so far were performed on stoichiometric NiAl. 
In general, very few studies on the effect of the local chemical composition on crack-tip processes exist \cite{Kermode2013,Liu2015b,Bitzek2015a}, and none of them in NiAl. 

Here we combine nanomechanical tests on different regions of compositionally graded pure NiAl single crystals with atomistic simulations to investigate the influence of Al concentration on the fracture toughness of B2 NiAl.

\begin{figure}[ht]
	\center
	\includegraphics[width=0.4\textwidth]{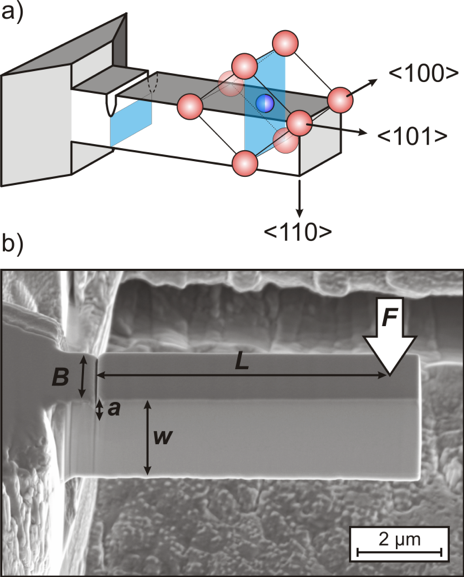} 
	\caption{\label{f1}
	a) Schematic of a microcantilever used to tests cracks in the “soft” orientation together with the NiAl unit cell and crystallographic directions. The crack plane is parallel to a $(110)$ plane. b) SE micrograph of the geometry of a microcantilever.
	}
\end{figure}

\section{Materials and Methods}

\subsection{Sample preparation and testing}
The influence of the chemical composition on the mechanical properties of the B2 phase was investigated using two single crystalline NiAl samples. Both were manufactured from pure elements with a purity of $> 99.95$\% for Ni and $>99.999$\% for Al. 
Details on the fabrication of the single-crystalline NiAl samples can be found in \cite{Bergmann1994,Bergmann1995,Tho99}. 
To introduce a gradient in chemical composition, sample 1 was annealed in a furnace under air atmosphere at $1100\, ^\circ$C for 100 h. 
The gradient, which developed due to the growth of alumina at the surface, was analyzed by energy dispersive X-ray diffraction. 
Sample 1 covers only the Ni-rich part of the phase field. 
Therefore, a second NiAl single crystal was aluminized in a pack cementation process and subsequently diffusion annealed  to increase the Al-content (sample 2). 
The samples were subsequently ground with an angle of $5^\circ$ towards the surface and finished with electropolishing. 
The microcantilevers were then prepared according to the routine proposed by Iqbal et al. \cite{Iqbal12} along the edge of the sample in different positions of the chemical gradient. 
The advantage of this procedure is that all microcantilevers have the same crystallographic orientation, which was confirmed by electron backscattered diffraction (EBSD) measurements. 
The orientation of the microcantilevers is shown in Fig.\ref{f1}\,a). 
Cracks with (101) plane orientation were introduced such that their crack front was oriented along the $\langle 110\rangle$ direction. 
This orientation is often referred to as “soft” orientation, because the yield stress in tensile test along this direction is below 270\,MPa \cite{Ebrahimi1998,Pascoe1968}, much lower than the one in the “hard” $\langle 100 \rangle$ direction ($\approx 1400$\,MPa) \cite{Pascoe1968}. 
The \{110\} planes were shown to be the natural cleavage planes of NiAl with a relatively low fracture toughness of 4\,MPa\,m$^{1/2}$\cite{Tho99}. 
The FIB milled microcantilevers were tested inside a scanning electron microscope (SEM) (Crossbeam 1540 Esb, Carl Zeiss AG, Germany) using a micromanipulator equipped with a force measurement system (FMS) (Kleindiek Nanotechnik, Germany).
This allows the in-situ observation of the experiment and also to track the displacement with digital image correlation (DIC).
Here, the software VEDDAC 5.1 \cite{VEDDAC} was used to evaluate the displacement fields. 
Figure \ref{f1}\,b) shows the final microcantilever geometry. 
In the following, the force that is applied by the indenter tip is denoted by $F$, the beam thickness by $B$, the width by $w$, the notch length by $a$ and the distance from the indenter tip to the notch by $L$. 
Macroscopic experiments to determine the fracture toughness make use of nominally atomically sharp cracks introduced by fatigue pre-cracking. 
This is not possible in the microscopic setup. 
Instead, a notch is milled from the top with a fine FIB current of 10\,pA and 30\,kV acceleration voltage. 
The notch radius is approximately 10\,nm, similar to the ones in \cite{Ast14}. 
For a more detailed description of the testing methodology the reader is referred to \cite{Iqbal12,Schaufler2012,Ast14}.

\subsection{Fracture mechanical analysis}
The fracture toughness can be calculated from a notched bending test according to Eq.~\ref{eq1}, if the plastic zone around the crack tip is small compared to the dimensions of the bending beam \cite{E399}:
\begin{equation}
\label{eq1}
K_\textrm{IQ}=\frac{F_\textrm{Q} L}{B W^\frac{3}{2}} f\left(\frac{a}{W}\right)
\end{equation}
This equation uses the maximum force $F_\textrm{max}$ for $F_\textrm{Q}$ at the point of fracture and the beam geometry as introduced above. 
A geometry function for this beam geometry, which only depends on a and w was determined by finite element modeling (FEM) by Iqbal et al. \cite{Iqbal12}. 

According to the standards for fracture toughness determination \cite{E399,E1820}, several requirements must be fulfilled. 
Due to the small size of the microcantilevers, not all stated requirements could be met in this setup. 
The present approach was however shown to lead to results comparable to macroscopic tests \cite{Iqbal12,Web2012,Ast14}. 
If, however, the extent of the plastic zone at the crack tip is comparable to the specimen size, and the load-displacement curve shows a deviation from the elastic behavior, elastic-plastic fracture mechanics (EPFM) has to be used instead of Eq.~\ref{eq1}.
This requires the calculation of the $J$-integral that takes the elastic $J_\textrm{el}$ and the plastic contribution $J_\textrm{pl}$ into account \cite{E1820}:
\begin{equation}
\label{eq2}
J=J_\textrm{el}+J_\textrm{pl}=\frac{K_\textrm{IQ}^2(1-\nu^2)}{E}+\frac{\eta A_\textrm{pl}}{B(W-a)}.
\end{equation}

Here $K_\textrm{IQ}$ is calculated with Eq.~\ref{eq1} using $F_\textrm{Q} = F_{0.95}$ and is then accordingly denoted by $K_\textrm{I,0.95}$. 
Following \cite{E1820}, $F_{0.95}$ is determined from the load-displacement plot as the force at the intersection of a straight line through the origin with a slope of $0.95E$ (see Fig.~\ref{f4}). 
The prefactor $(1-\nu^2)/E$ in Eq.~\ref{eq2}, which combines the Young's modulus $E$ and Poisson ratio $\nu$, is calculated from the elastic constants $c_\textrm{ij}$ for the specific chemical composition \cite{Rusovic1977} and the orientation of the crack following \cite{Sih1968}. 
The plastic contribution $J_\textrm{pl}$ is determined by using the area $A_\textrm{pl}$ under the load-displacement curve, excluding the elastic contribution. 
The value of the constant  is set to 2. 
The $J$-integral is then used with the expression
\begin{equation}
\label{eq3}
K_{\textrm{Q},J}=\sqrt{\frac{JE}{1-\nu^2}}
\end{equation}
to calculate the fracture toughness $K_{\textrm{Q},J}$. 
As the samples do not fulfil all the requirements of the standards for fracture toughness determination, $K_{\textrm{Q},J}$ is referred to as conditional fracture toughness \cite{Wurster2012}.

\subsection{Calculation of theoretical fracture toughness}
According to Griffith \cite{Griffith1921}, the theoretical plane strain fracture toughness $K_\textrm{G}$ can be expressed in terms of the energy release rate $G$ by
\begin{equation}
\label{eq4}
K_{\textrm{G}}=\sqrt{G_\textrm{G}B^{-1}},
\end{equation}
with $B$ being an orientation dependent elastic constant that can be expressed in terms of $c_{ij}$ \cite{Sih1968}. For fracture to take place, the critical energy release rate $G_\textrm{G}$ has to equal the energy of the surfaces $2\gamma_s$ created by the propagating crack:
\begin{equation}
\label{eq5}
G_{\textrm{G}}=2\gamma_s.
\end{equation}

To evaluate $K_{\textrm{G}}$ for different chemical compositions, $c_{ij}$ and $\gamma_s$ need to be determined for these compositions. 
This was achieved by performing MS calculations on a sample containing $15\times 15 \times 15$ B2 NiAl unit cells. 
The respective off-stoichiometric compositions were modeled by randomly replacing Ni atoms by vacancies for Al-rich compositions, and Al atoms by Ni atoms on the Ni-rich side \cite{Meyer2001}. 
For the calculation, periodic boundary conditions (PBC) were used in all directions, and the box size was adjusted to obtain a stress-free minimum energy configuration. The surface energies were calculated by performing the MS simulations without PBC in the $[110]$ direction. 
The calculations were performed on three different samples to provide an average value for the elastic constants and surface energies, independent of the individual atomic arrangement. 
All MS simulations were performed with the FIRE algorithm \cite{Bitzek2006}, and the EAM potential for NiAl by Pun and Mishin \cite{Pun2009a} was used in all simulations. 
This potential was shown to reproduce well the equilibrium and defect properties of NiAl as determined from ab-initio simulations and experiments \cite{Pun2009a}. 
The relevant properties of this potential are provided in Tab.~\ref{t1}.

\begin{table}[]
  \begin{center}
    \begin{tabular}{lccc}
    \hline
    & Ni$_{55}$Al$_{45}$ &NiAl &Ni$_{45}$Al$_{55}$\\
    \hline
    $a_0$ (nm)&0.2824&0.2832&0.2819\\
$c_{11}$ (GPa)&197&191&199\\
$c_{12}$ (GPa)&153&143&126\\
$c_{44}$  (GPa)&124&122&98\\
$E_{[110]}$ (GPa)&155&160&180\\
$A$&6.6&5.1&4.9\\
$\nu_{\langle 100 \rangle}$&0.44&0.43&0.39\\
$\gamma_{(110)}$ (Jm$^{-2}$)&1.94&1.89&1.73\\
$\gamma_{\textrm{usf}}$ (Jm$^{-2}$)&1.63&1.56&1.40$\pm0.06$\\ 

\hline
    \end{tabular}
    \end{center}
    \caption{Overview of fracture-relevant properties as determined by atomistic simulations using the NiAl-potential of Pun et al. \cite{Pun2009a} for varying Al-concentrations: lattice constant $a_0$, elastic constants $c_{ij}$, Young’s modulus in $[110]$ direction $E_{[110]}$, anisotropy ratio $A$, Poisson’s ratio in cube directions $\nu_{\langle 100 \rangle}$, surface energy of $\{110\}$ surfaces $\gamma_{(110)}$, unstable stacking fault energy in $[001]$ direction on $(100)$ planes $\gamma_{\textrm{usf}}$. For the non-stoichiometric concentrations, the calculations are repeated three times to reduce the statistical error, which lies within the given precision except where stated otherwise.}
    \label{t1} 
\end{table}

\begin{figure*}[th]
	\center
	\includegraphics[width=0.7\textwidth]{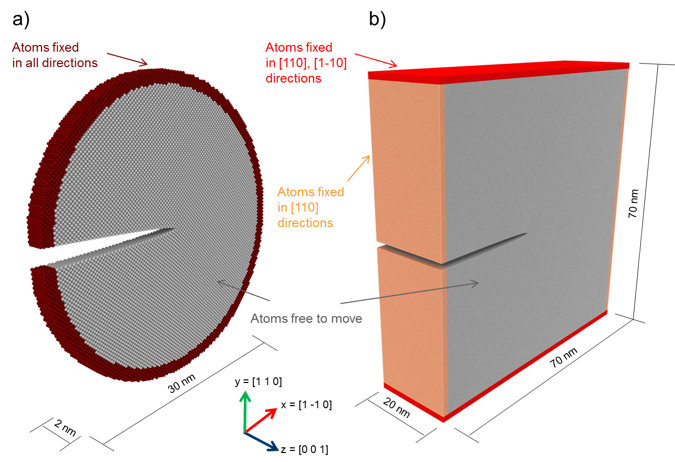} 
	\caption{\label{f2}
	Simulation setup for a) determination of fracture toughness and b) simulation of propagating cracks. Periodic boundary conditions along the crack front ($z$)-direction are used in both cases.
	}
\end{figure*}

\subsection{Atomistic fracture simulations}

The simulation setup for the determination of the fracture toughness for cracks under mode-I loading is illustrated in Fig.~\ref{f2}\,a). 
An atomically sharp crack is introduced in a cylindrical sample of radius $R=15$~nm by displacing the atoms according to the linear elastic solution of a crack in an infinite anisotropic elastic body \cite{Sih1968,Moller2013}. 
PBCs are used in crack front ($[001]$) direction and atoms in a boundary layer of two times the cut-off radius of the potential are fixed. 
To determine the fracture toughness, MS calculations are performed where the applied stress intensity is iteratively increased in steps of $\Delta K_I=0.007$~MPa~m$^{\frac{1}{2}}$. 
The critical stress intensity factor $K_{Ic}$ is taken as the stress intensity at which the entire crack propagated by at least one atomic distance. 
For details of the simulation procedure see \cite{Moe14msmse,Moe14am}.  The calculations were performed on five samples to study the influence of the statistical distribution of atomic species in the off-stoichiometric samples.
Doubling the sample size to $R=30$~nm only lead to a change in the fracture toughness within one step size $\Delta K_I$.

A different set-up (Fig.~\ref{f2}\,b) is used to study propagating cracks. 
A sample size of $70\times70\times20$~nm with PBC in the $[001]$ direction and fixed boundary conditions in the other directions is used. 
A blunted crack is introduced at the center of the sample by removal of one half-plane of atoms. 
Afterwards, the displacement field of a semi-elliptical crack is applied and the configuration is strained by a value slightly below the critical strain $\epsilon_G$ according to the Griffith energy balance as determined following \cite{Bit13}. The crack tip is located at 40\% of the box length in $[\bar1 1 0]$ direction. 
The entire configuration is relaxed ensuring zero stress in PBC direction. 
Standard MD simulations with a starting temperature of 0~K are performed while the sample is homogeneously strained in $[110]$ direction at a strain rate of $10^8$~s$^{-1}$, while maintaining zero stress in the orthogonal directions. Defect analysis is carried out using slip vector analysis method \cite{Zimmerman2001}. Open Visualization Tool (OVITO) is used for visualization of the atomistic simulation data \cite{Stukowski2009}.

%Visualization and defect analysis are carried out with Atomviewer \cite{Beg12,atomviewer}.

\section{Results}

\subsection{Microstructural and chemical analyses}
\begin{figure}[]
	\center
	\includegraphics[width=0.4\textwidth]{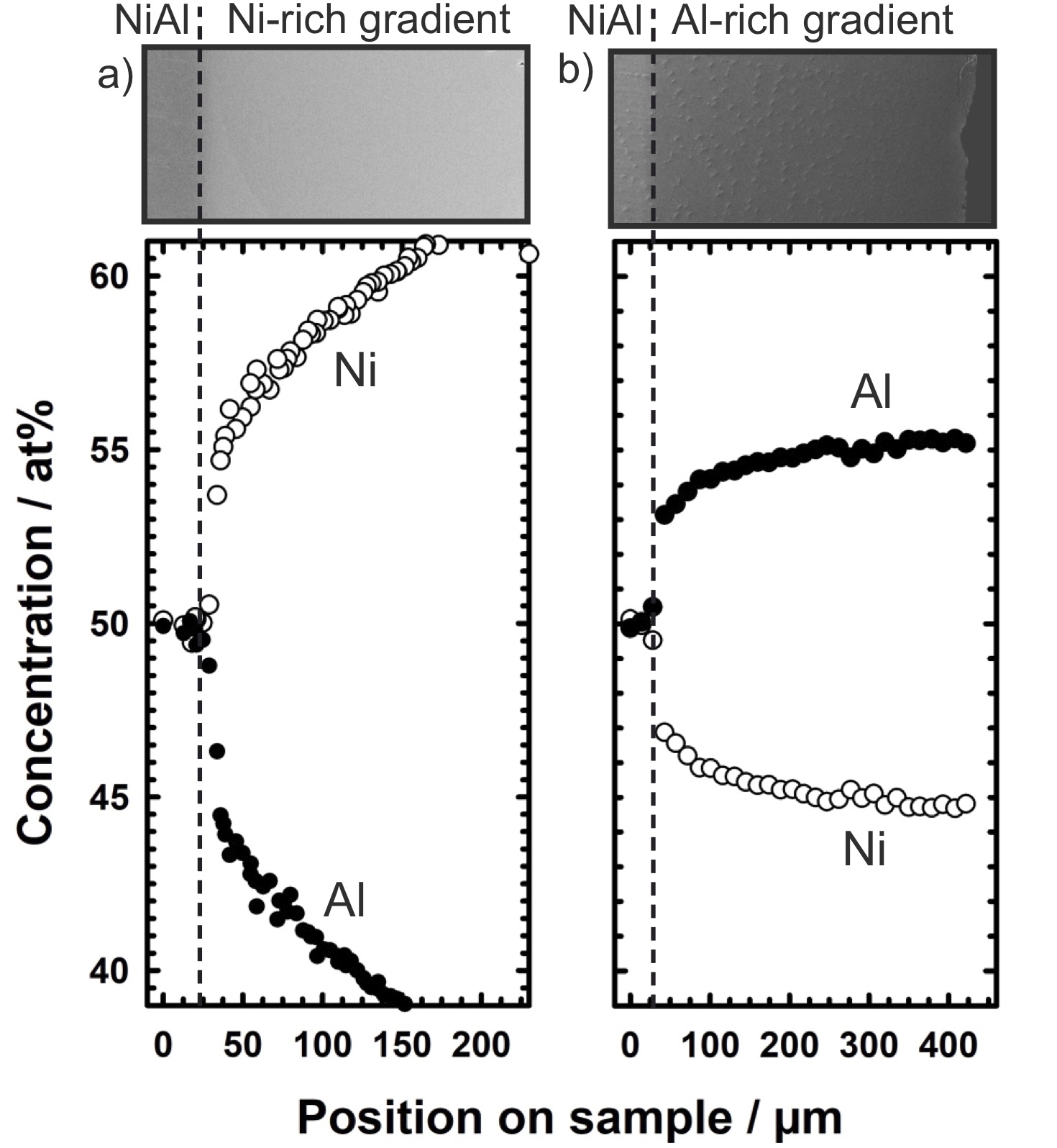} 
	\caption{\label{f3}
	a) Microstructures and corresponding concentration profiles of both samples with gradients towards high a) Ni- and b) Al-content.
	}
\end{figure}

The micrographs in backscattered electron contrast and the chemical composition of both B2 NiAl samples are shown in Fig.~\ref{f3}.

With increasing Ni- and Al- content towards the right side of the samples a light grey and dark grey zone with the chemical gradient can be distinguished from the grey stoichiometric NiAl on the left side of both samples. The Al-content of sample 1 (see Fig.~\ref{f3}\,a) decreases sharply within a distance of 10 $\mu$m from the stoichiometric composition to 45 at\% and then gradually falls off to a concentration below 40 at\%. The overall gradient in the chemical composition of sample 2 (see Fig.~\ref{f3}\,b) is less steep than in sample 1, but also exhibits a jump in the concentration profile close to the stoichiometric composition.

\begin{figure}[]
	\center
	\includegraphics[width=0.4\textwidth]{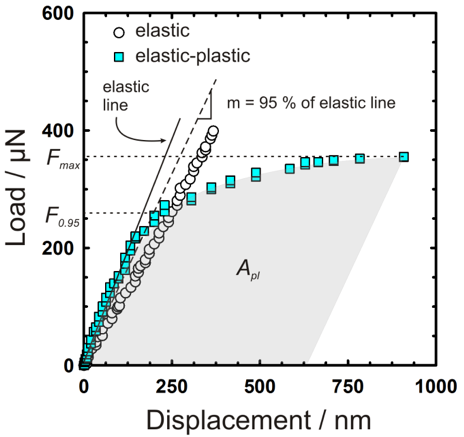} 
	\caption{\label{f4}
	Representative force-displacement curves for elastic and elastic-plastic fracture behavior together with the constructions to determine the parameters in Eq.~\ref{eq2}.
	}
\end{figure}

\subsection{In-situ microcantilever bending experiments}
\begin{figure}[]
	\center
	\includegraphics[width=0.4\textwidth]{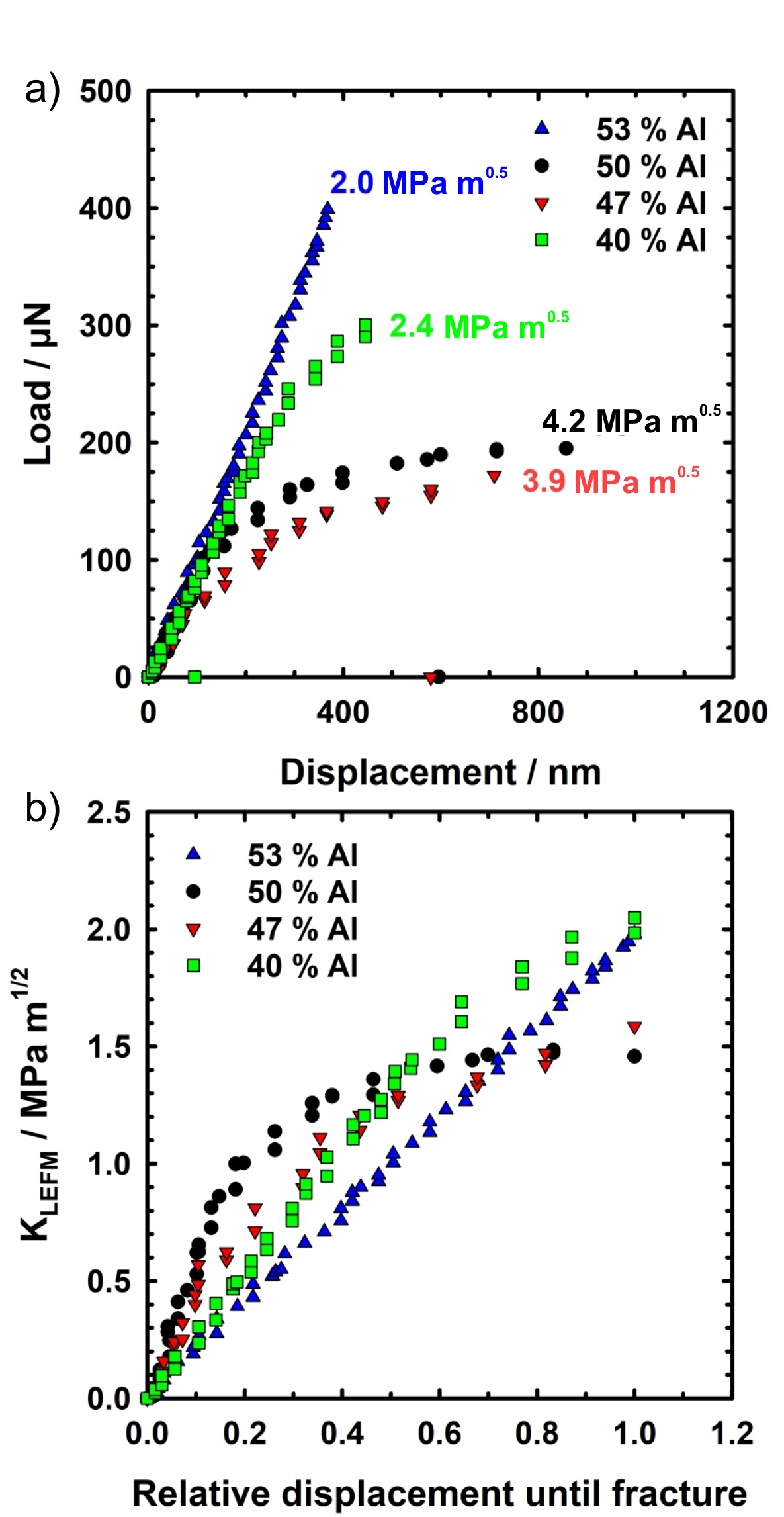}
	\caption{\label{f5}
	%\todo{SN: add units of K in a)}
	(a) representative load-displacement curves for different Al-concentrations. (b) Plot of stress in-tensity factor vs. relative displacement as calculated from (a) using Eq.~\ref{eq1}. The fracture toughness values determined by Eqs.~\ref{eq1} and \ref{eq2} are provided next to the data points in (a). 	}
\end{figure}

Depending on the composition, purely linear elastic as well as elastic-plastic load-displacement response of the notched microcantilevers could be observed during the bending experiments. 
Figure \ref{f4} shows two representative cases together with the constructions to determine the necessary parameters to calculate the fracture toughness according to Eq.~\ref{eq1} or \ref{eq2}. 
Representative load-displacement curves for different compositions can be found in Fig.~\ref{f5}\,a). 
The linear elastic parts do not have exactly the same slope, because the distance $L$ of the crack from the loading point, i.e., the contact with the indenter, is not identical in all cases. 
The different $L$ are however considered in the determination of the fracture toughness, Eqs \ref{eq1} and \ref{eq3}, see also Fig.~\ref{f5}b). 
The Al-rich microcantilevers show a linear-elastic behavior up to the point of fracture. Here, the maximum load $F_\textrm{max}$ can be used for the value of $F_Q$ in Eq.~\ref{eq1} to determine the fracture toughness.  
That way, the average fracture toughness of cantilevers with Al-rich compositions was determined to $K_{Ic}$ of $2.1 \pm 0.2$ MPa~m$^{1/2}$. 
Stoichiometric and Ni-rich NiAl microcantilevers, however, show an elastic-plastic behavior. 
For these compositions, the condition $F_\textrm{max}/F_{0.95} > 1.1$ \cite{E1820} is not met, and EPFM (Eqs.~\ref{eq2} and \ref{eq3}) rather than LEFM (Eq.~\ref{eq1}) has to be used. 
The plastic part of the J-integral is proportional to $A_\textrm{pl}$, the area under the load-displacement curve minus the elastic contribution \cite{E1820}. 

\begin{figure}[]
	\center
	\includegraphics[width=0.4\textwidth]{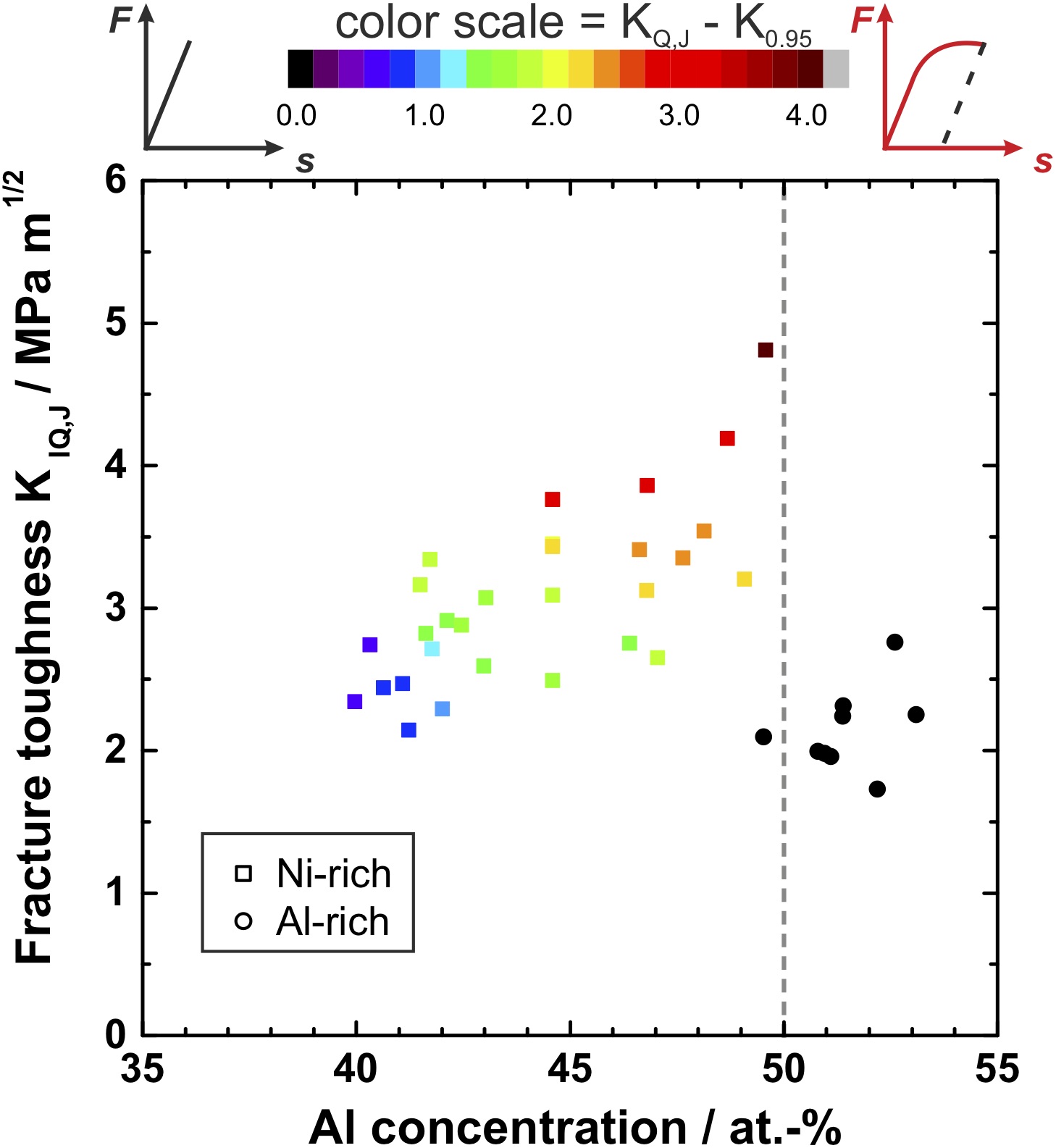} 
	\caption{\label{f6}
	Experimentally determined NiAl fracture toughness in dependence of Al concentration. The color code of the Ni-rich symbols (squares) represents the contribution of the plastic deformation to $K_{Q,J}$, as shown by the schematic load-displacement ($F-s$) curves for linear-elastic (left) and elastic-plastic (right) fracture behavior. Dashed lines indicate the range below which the B2 lattice structure changes to L10 martensite and the stoichiometric NiAl composition. 	}
\end{figure}

Figure \ref{f5} a) clearly shows that for the Ni-rich compositions, $A_\textrm{pl}$ decreases with decreasing Al-content. 
The fracture toughness values of all tested specimens are shown in Fig.~\ref{f6}. 
For purely brittle materials, Eqs.~\ref{eq1} and \ref{eq3} lead to the same fracture toughness $K_{IQ}= K_{Q,J}$. 
Therefore, only $K_{Q,J}$ is plotted in Fig.~\ref{f6}. 
In the case fracture is accompanied by plastic deformation, the difference $K_{Q,J} - K_{IQ}$ can be used to evaluate the amount of plasticity and the deviation from the purely elastic behavior of the load-displacement curve. 
The data points in Fig.~\ref{f6} are color-coded according to this measure. 
In this Figure, the increase of plastic deformation and fracture toughness with increasing Al concentration in the Ni-rich part of the B2 phase is clearly evident. 
In Al rich NiAl, however, no plastic contribution to $K_{Q,J}$ can be discerned, and the fracture toughness appears not to be influenced by the Al concentration.  
In the B2 part of sample 1, the microcantilever with composition closest to the stoichiometric composition has the highest fracture toughness ($K_{Q,J}\approx 4.7$ MPa~m$^{1/2}$) and the largest deviation from the LEFM solution  
($K_{IQ}\approx1.7???$~MPa~m$^{1/2}$). 
%In the areas where the Al-concentration is so low that L1$_0$ martensite developed, the fracture toughness shows a large scatter with values between 2-5 MPa~m$^{1/2}$.

\begin{figure}[]
	\center
	\includegraphics[width=0.45\textwidth]{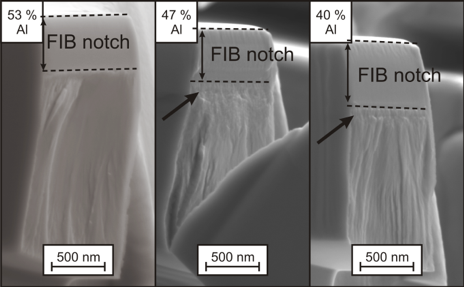} 
	\caption{\label{f7}
Fracture surface of microcantilevers with different chemical composition given in at\%. The arrows indicate where the initially stable crack growth changed to unstable crack propagation. 
	}
\end{figure}

\begin{figure}[]
	\center
	\includegraphics[width=0.47\textwidth]{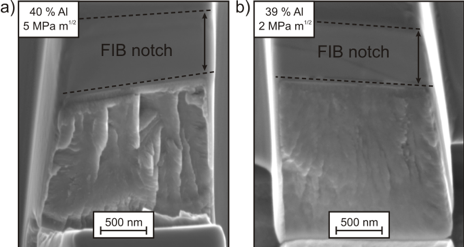} 
	\caption{\label{f8}
	Fracture surfaces of microcantilevers with almost identical Al-content but large difference in fracture toughness of a) 5\MPam and b) 2\MPam.
	}
\end{figure}

Fracture surfaces of broken microcantilevers with different Al concentrations are shown in Fig.~\ref{f7}.
Al-rich NiAl like the example at 53 at\% in Fig.~\ref{f7} shows typical cleavage fracture with a smooth crack onset right below the notch. 
Cracks in Ni-rich B2 NiAl, however, show a more ragged appearance, see Figs.~\ref{f7} and \ref{f8}. 
%% I would not show data of cantilevers with a composition below 40% Al. I would delete the following two sentences.
%Cantilevers with similar chemical compositions can show differences in fracture toughness up to about 1 MPa~m$^{1/2}$ as seen in Fig.~\ref{8}. 
%In these cases, the roughness of the fracture surface is directly correlated with the fracture toughness, see Fig.~\ref{10}.  
A common feature for all cracks in Ni-rich microcantilevers is a short initial phase of stable crack growth before final fracture. 
The arrow in Fig.~\ref{f7} shows where the final fracture initiated. 
The step-like structure below and parallel to the notch suggests that the crack might have momentarily stopped before finally fracturing the cantilever. 
The initial stable crack growth can also be seen in movie M1 in the supplementary material \ref{M1}.

\subsection{Atomistic calculations}
The lattice constant $a_0$, the elastic constants $c_{ij}$ and the surface energy of ${110}$ surfaces $\gamma_{(110)}$ have been calculated for B2 NiAl for the entire composition range between 35 at\% Al to 60 at\% Al. 
Values for 45, 50, and 55 at\% Al are provided in Tab.~\ref{t1}. 
While $c_{11}$ is lowest for stoichiometric NiAl, $c_{12}$ and $c_{44}$ show a significant decrease with increased Al concentration. 
It is interesting to note that the Young’s modulus in $[110]$ direction $E_{[110]}$, however, shows an increase with increasing Al content, see Tab.~\ref{t1}. 
This can be understood by calculating the anisotropy ratio $A$, which also shows a decline with increasing Al concentration. With decreasing elastic anisotropy, the differences between the Young’s moduli in different directions have to become smaller, i.e., $E_{[111]}$ decreases while $E_{[110]}$ and $E_{[100]}$ increase. 

\begin{figure}[]
	\center
	\includegraphics[width=0.5\textwidth]{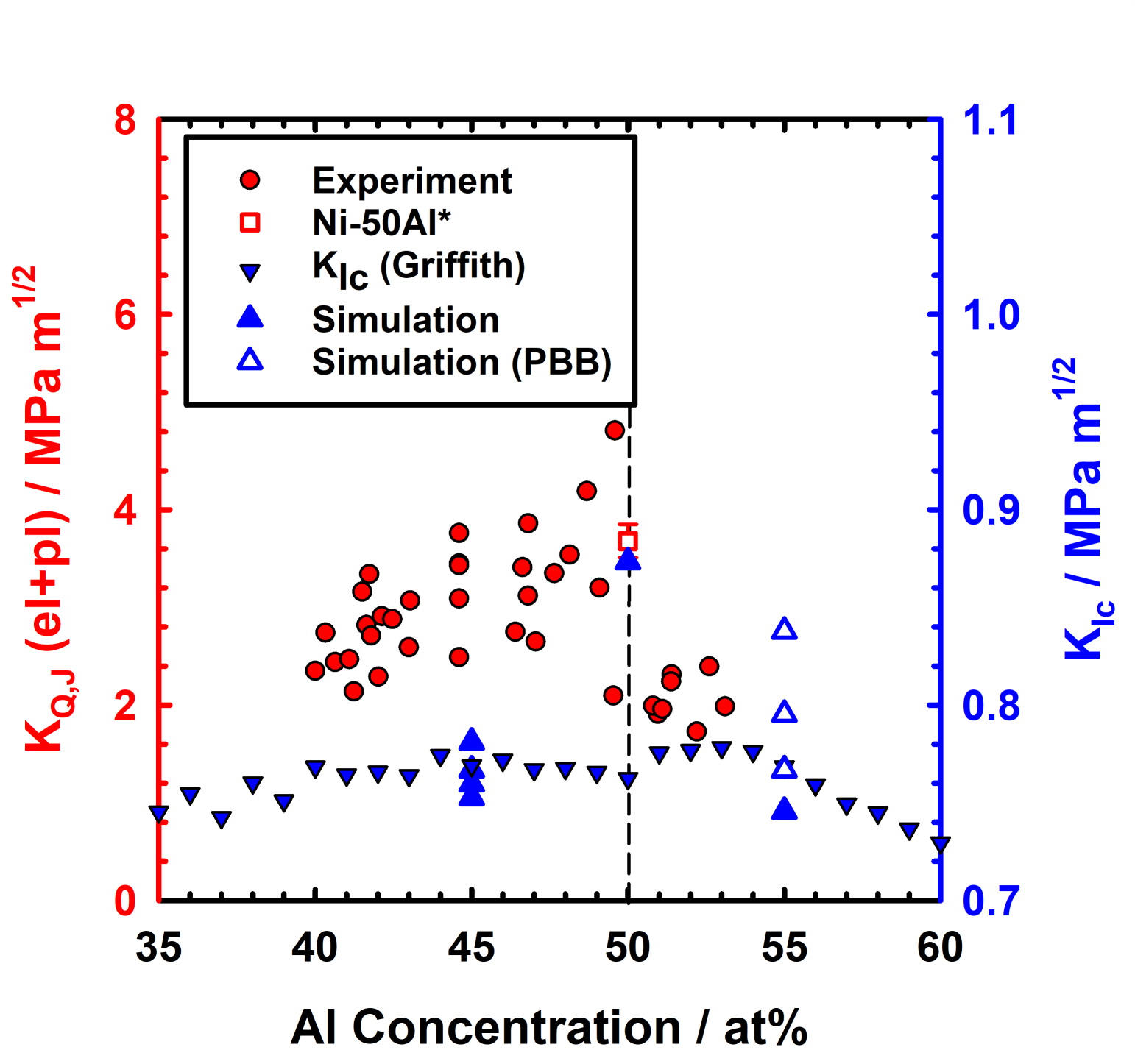} 
	\caption{\label{f9}
	Experimentally determined fracture toughness (identical to Fig.~\ref{f6}) together with the atomistically calculated fracture toughness following Griffith (Eq. (4)) KG and the results of the quasistatic simulations. Ni-50Al* represents the results from Ast et al \cite{}. Please note the different scales for experimental data (red symbols, left axis) and atomistic calculations (blue symbols, right axis).
	}
\end{figure}

Based on the determined values of the surface energy and the elastic properties, the theoretical value for the critical stress intensity factor $K_G$ can then be calculated following the Griffith concept \cite{Griffith1921} taking into account the anisotropic relationship between the applied $K_I$ and the crack tip displacement field \cite{Sih1968}. 
This theoretical fracture toughness according to LEFM assumes perfectly brittle fracture and is the lowest possible fracture toughness of this material. 
Its dependence on the composition is shown in Fig.~\ref{f9}. 
Its value is nearly constant in the Ni-rich part ($K_G\approx 0.76$~MPa~m$^{1/2}$), slightly increases in the Al-rich part close to the stoichiometric composition and then decreases with increasing Al concentration. 

\begin{figure}[]
	\center
	\includegraphics[width=0.48\textwidth]{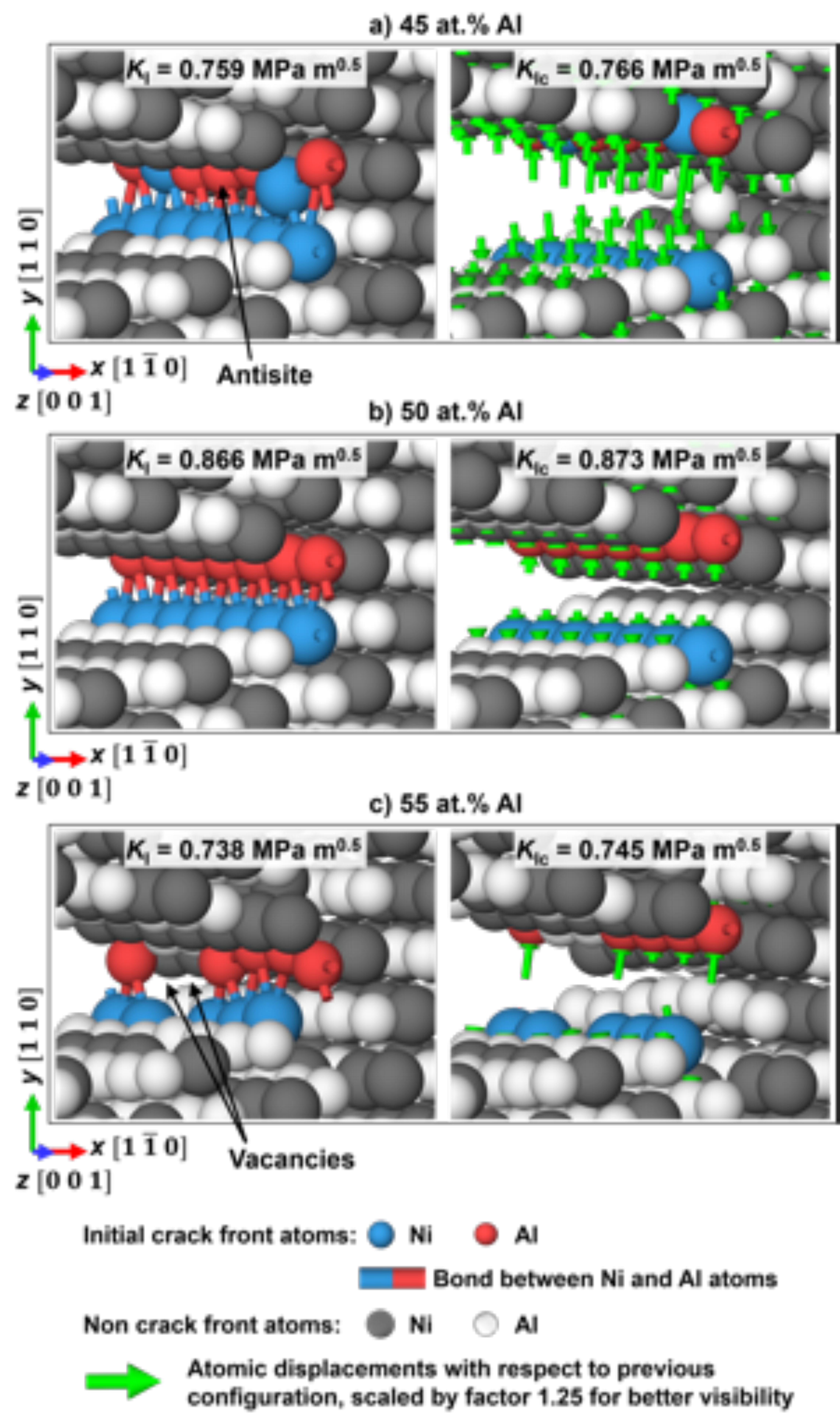} 
	\caption{\label{f11}
	Example for the determination of the critical stress intensity factor $\left( K_{Ic} \right) $. In all cases the entire crack propagated by one atomic distance within one load increment, resulting in a unique value for $K_{Ic}$ .
	}
\end{figure}

\begin{figure*}[]
	\center
	\includegraphics[width=0.975\textwidth]{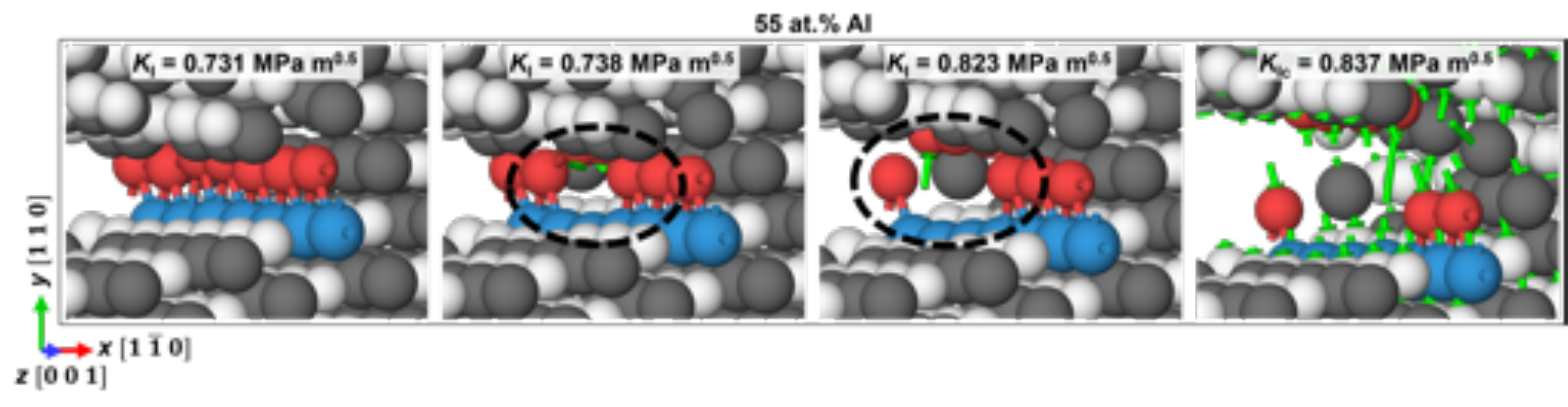} 
	\caption{\label{f12}
	Example for the determination of the critical stress intensity factor  when prior bond breaking (PBB) processes take place before the entire crack has advanced by one atomic distance at the load $K_{Ic}$. This behavior was only observed when structural vacancies were present close to the crack tips in 
 Ni- 55 at.$\%$Al. PBB are highlighted by dashed ellipses. The atom colours and green arrows have same meaning as in figure \ref{f11}.
	}
\end{figure*}

Whereas the determination of $K_{Ic}$ by static simulations is well-established for uniform or ordered structures, determining the fracture toughness for crystals with atomic-scale disorder like in the off-stoichiometric cases is less straight-forward. 
In general, a crack can propagate by breaking all crack-tip bonds 
within one load increment, or by subsequently breaking of individual or multiple bonds during successive load increments.
The first case is exemplary shown in Fig.~\ref{f11}. 
It is obviously the case for stoichiometric NiAl, where all crack-tip bonds are identical, see Fig. \ref{f11}b.
However, also all cracks in Al-rich NiAl propagated within one load increment, see e.g. Fig. \ref{f11}a.
In contrast all but one crack in Ni-55 at.$\%$Al showed prior bond breaking (PBB) processes before the crack propagated at $K_{Ic}$, see Figs.~\ref{f11}c and \ref{f12}.
The so determined values of $K_{Ic}$ are shown in Fig.~\ref{f9} and provided in Tab.~\ref{t2}.
Interestingly, PBB in the Al-rich samples was only observed when the vacancies were not located directly at the crack front and these samples also showed higher fracture toughness values than the sample with vacancies directly at the crack tip that showed no PBB. 
None of the simulated samples showed any sign of 
crack-tip plasticity, see the exemplary figures in the supplementary Material, Figs.~\ref{S1}-\ref{S4}. 

It is furthermore interesting to note that in Fig.~\ref{f9}  $K_{Ic}$ 
can be smaller than predicted by the Griffith equation. 
This is the case as the propagating crack only produces a surface increment of one atomic spacing, whereas the surface energy used to calculate $K_{Ic}$  according to Griffith in Fig.~\ref{f9} was averaged over a large area as well as multiple realizations. 
The present, established method therefore can not be used to determine thermodynamically reasonable values for the fracture toughness in atomically disordered systems. This would require 
the propagation of the crack over large distances, which -- under pure $K$-controlled loading --  is not possible with the currently available boundary conditions.

\begin{figure*}[]
	\center
	\includegraphics[width=\textwidth]{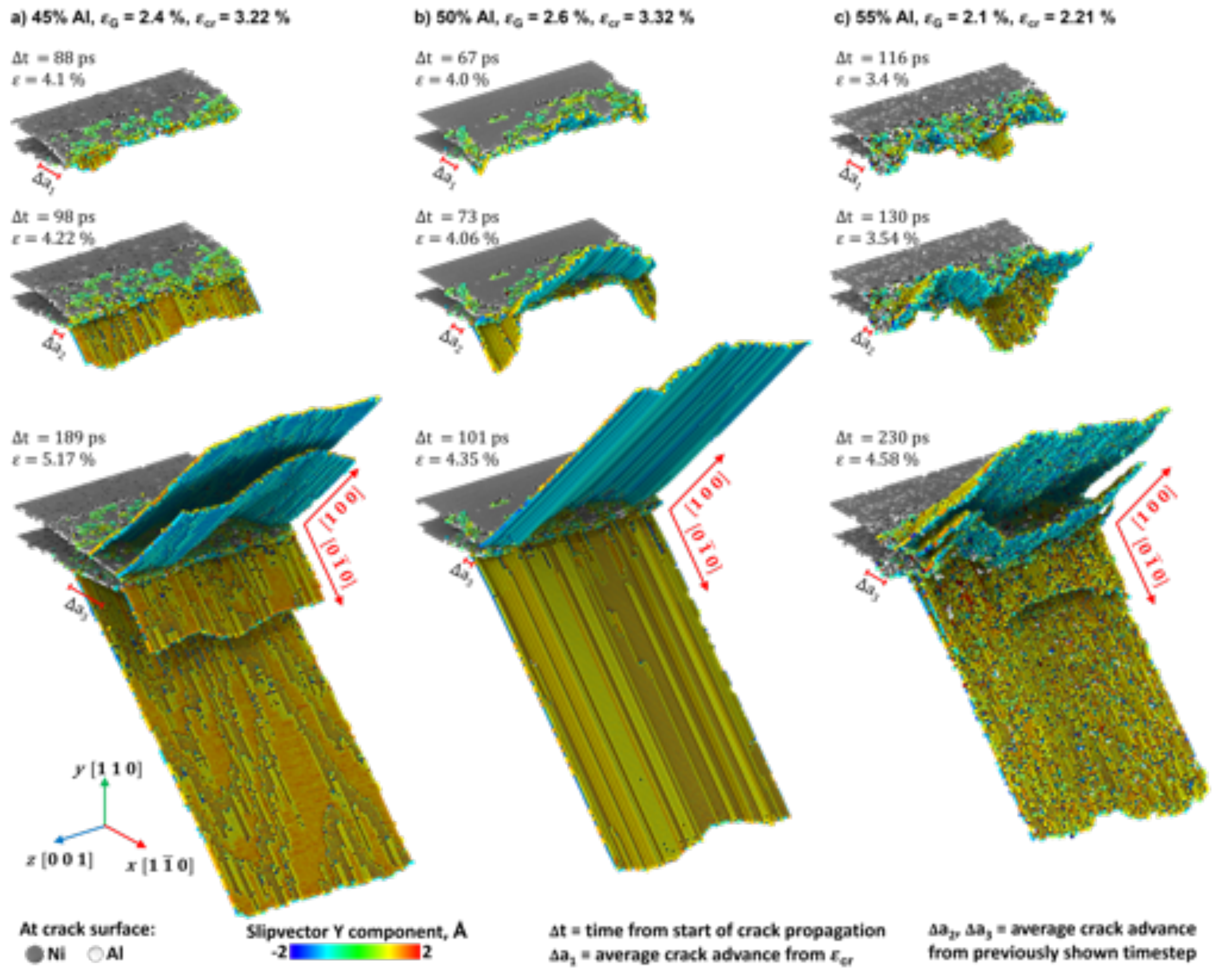} 
	\caption{\label{f10}
	Snapshots from the MD simulations of cracks in NiAl samples of different compositions subjected to a constant strain rate $\epsilon_{yy}=10^8$~s$^{-1}$:  during the process of dislocation emission (top row), propagation of crack and dislocations (middle row), and emission of further dislocations (bottom row). Only atoms belonging to defects other than the initial crack (dark grey: Ni, light gray: Al) are shown using slip vector analysis. The arrows shown next to the dislocations in the bottom row indicate the Burgers vector directions of the dislocations.
	}
\end{figure*}

Crack propagation therefore was studied using a strain- rather than a $K$-controlled setup.  
 Snapshots from the MD simulations are shown in Fig.~\ref{f10}. 
For the stoichiometric sample, Fig.~\ref{f10}b), the crack propagates for a couple of lattice constants and then starts to emit a dislocation on the $(010)[100]$ slip system.
The crack is slowed down by the emission of the dislocation, but under the constant strain rate of $10^8$~s$^{-1}$, the crack propagates further and emits another dislocation on the mirror symmetric $(100) [0 \bar 1 1]$ slip system. 
The entire process can be seen in supplementary movie M2 \cite{}.  
The sample with 45 at\% Al, Fig.~\ref{f10}a), behaves similarly to the stoichiometric sample: The Griffith strains $\epsilon_G$ at which the elastically stored energy in the sample is identical to the surface energy are comparable, as is the critical strain $\epsilon_\textrm{cr}$ at which the crack starts to propagate. 
Also the emission of the first dislocation (on the $(100) [0 \bar 1 1]$ system) takes place at roughly the same strain, see also Tab.~\ref{t2}. 
The emission of the second dislocation, however, takes place significantly later compared to the stoichiometric sample, see supplementary movie M3 \cite{}. 
The Griffith strain for the sample with 55 at\% Al, Fig.~\ref{f10}c), is smaller than for the other samples (Tab.~\ref{t2}). 
More importantly, however, is that crack propagation sets in at a value of  $\epsilon_\textrm{cr}$ that is only slightly larger than $\epsilon_G$ and 50\% lower compared to the stoichiometric and Ni-rich samples. 
In addition, the crack front appears rougher and dislocation emission starts on both slip systems. 
Furthermore, the propagating dislocations create significant debris, see also supplementary movie M4 \cite{}.

\begin{table}[]
  \begin{center}
    \begin{tabular}{lccc}
    \hline
    &  Ni$_{55}$Al$_{45}$ &NiAl &Ni$_{45}$Al$_{55}$\\
    \hline
    Quasistatic:& &&\\
    $K_G$ (\MPam) & 0.769 & 0.762 & 0.768\\
    $K_{Ic}$ (\MPam) & 0.77 $\pm$0.02 & 0.87 & 0.79$\pm$0.05\\
    behavior: & B & B & B\\
  \hline
Dynamic:& &&\\
$\epsilon_G$ (\%) &2.4&2.6&2.1\\
$\epsilon_{cr}$ (\%) &3.22&3.32&2.21\\
$\epsilon_\textrm{dislo}$ (\%) &4.1&4.0&3.4\\
behavior: & B+D & B+D & B+D\\
\hline
    \end{tabular}
    \end{center}
    \caption{Results of the quasistatic and dynamic fracture simulations: $K_G$  – theoretical fracture toughness according to Eq. (4); $K_{Ic}$ - critical applied stress intensity at which the crack either propagates or crack tip plasticity occurs; $\epsilon_G$ - critical strain according to the Griffith energy balance; $\epsilon_{cr}$ - critical strain at which large-scale bond breaking processes occur; $\epsilon_\textrm{dislo}$ – strain at which the first dislocation is emitted; Behavior: D – dislocation emission, B – brittle crack propagation.}
    \label{t2} 
\end{table}

\begin{figure*}[]
	\center
	\includegraphics[width=0.9\textwidth]{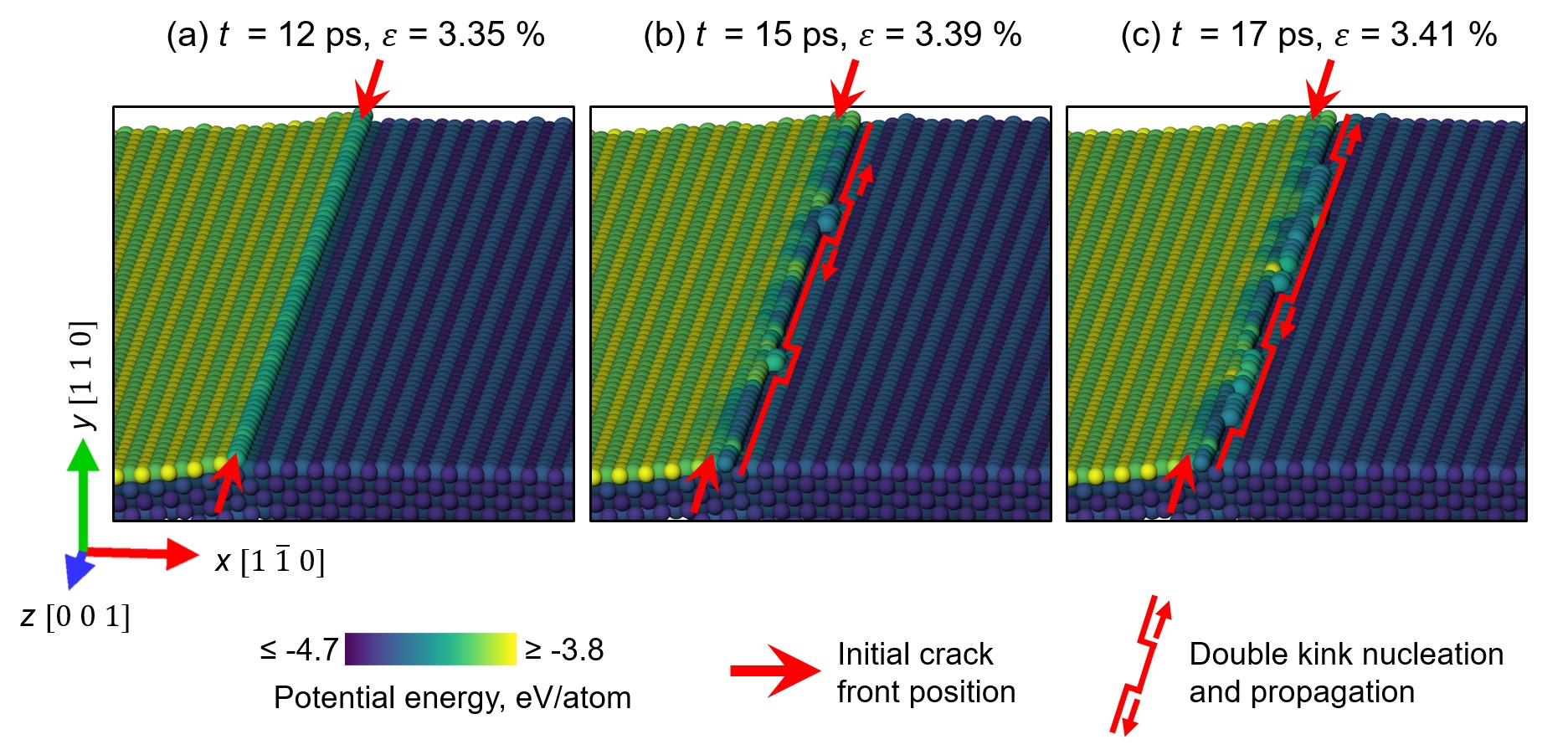} 
	\caption{\label{f14}
	Crack propagation by double kink nucleation and kink propagation for Ni- 50 at.$\%$ Al. Crack propagates along $x$ direction. Only the lower half of the sample is shown with atoms colored by their potential energy, clearly showing the fracture surface.
	}
\end{figure*}

A detailed analysis of the stoichiometric sample with high time resolution shows that event at 0K the crack propagates by the nucleation and migration of kink pairs, see Fig.~\ref{f14}.

\section{Discussion}

Macroscopic fracture experiments on stoichiometric B2 NiAl report fracture toughness values of the hard orientation between 5 and 9 MPa~m$^{1/2}$ \cite{Bergmann94,Tho99,Chang1992}. 
The recent work by Ast et al. showed that these values could well be reproduced in microscale experiments using the here-employed methodology \cite{Ast14}. 
Their values match well with our results on near-stoichiometric, Ni-rich NiAl, see Fig.~\ref{f9}. 
The fracture toughness of Al-rich samples ranges between 1.7 and 2.4 MPa~m$^{1/2}$, without any noticeable dependence on the Al concentration. 
This compares well with the Al concentration independent fracture toughness values of 2.2-2.9 MPa~m$^{1/2}$ determined by Wellner et al. \cite{Wellner2004} for polycrystalline NiAl films with 50.4-52.4 at.\% Al. 
For deviations from the stoichiometric compositions to the Ni-rich side, however, a nearly linear decrease with increasing Ni contents can be seen up to Ni concentrations of about 40 at\%.  
%For Ni-fractions higher than 62 at\%, stress induced phase changes from the B2 into the L1$_0$ structure were reported at crack tips \cite{18, 31}. 
%We therefore attribute the large scatter in the experiments at Ni concentrations below 40 at\% to the local (reversible) formation of martensite. 

The results from the quasistatic fracture simulations on the stoichiometric sample agree with the work by Ludwig and Gumbsch \cite{Ludwig1998} in so far that no dislocation emission was observed. 
The difference in their value for $K_{Ic}= 0.65$ MPa~m$^{1/2}$ to our result of 
$K_{Ic}= 0.87$ MPa~m$^{1/2}$ is most probably rooted in the different potential used in this study. 
The generally large differences between atomistically and experimentally determined values of the fracture toughness are well-known and are attributed to the idealized simulation setup, that precludes effects due to the activity of preexisting dislocations, imperfect cleavage surfaces or atomic-scale crack tip blunting, which are all present under typical experimental conditions. 
It is important to note, that only the stoichiometric composition showed significant lattice trapping \cite{Thomson1971a,Bitzek2015a}, which explains the difference between $K_G$ and $K_{Ic}$ in Tab.~\ref{t2}. 
The disordered crack front in the non-stoichiometric compositions leads to local changes in the bonding strength of the crack tip atoms and thus the possibility of kink formation and local conditions which favor {\em local} crack advance for values of $K_{Ic}$ smaller than the thermodynamically averaged $K_G$. 
The determination of a static fracture toughness from atomistic simulations of media with atomic-scale disorder thus requires statistical approaches \cite{Ponson2016}, which have not yet been applied in the context of atomistic fracture simulations.  

Initially, the MD simulation show the same behavior than the static simulations, namely brittle crack advance, see Fig.~\ref{f10}. 
However, after a short distance, the emission of $\langle 100 \rangle$ dislocations on $\{010\}$ planes sets in.  
That dislocations were not observed in the static simulations can have multiple reasons. 
The large strains necessary in the clamped strip geometry could have lowered the energy barrier to nucleate dislocations through the well-known tension-shear coupling \cite{Sun1993}. 
Dislocation emission could also be related to a dynamic crack tip instability, or the kinks and jogs in the propagating crack front provided preferable sites for dislocation nucleation \cite{Gor09}. 
In general it has been shown, that the emission of loops rather than straight dislocation lines as enforced by the boundary conditions and the static minimization in the setup Fig.~\ref{f2}a) requires lower activation energy \cite{Zhu2004c}. 
It is however important to note, that same types of dislocations than in the simulations were observed to have produced large slip traces in front of cracks in the hard orientation \cite{Ast14}.

Traditionally, the fracture toughness is seen as a result of the competition between dislocation nucleation and motion on the one hand and bond breaking processes at the crack tip on the other hand \cite{Bitzek2015a}. 
The concentration dependence of the experimentally determined fracture toughness in Ni-rich NiAl samples can be nicely correlated with the mobility of dislocations as evidenced by the concentration-dependence of the hardness, see Fig.~\ref{fxx5}a). 
Here, it is important to note that while the hardness averages over $\{110\}$ and $\{100\}$ dislocations, only the shielding dislocations on the $\{100\}$ planes, which show lower hardening coefficients \cite{Gumbsch1999} are relevant, see Fig.~\ref{f6} and \cite{Ast14}. 

This is however not the case for Al-rich NiAl. 
Even an increase of only 0.4 at\% Al from the stoichiometric composition leads to drastically decreased fracture toughness of NiAl \cite{Wellner2004}, which then stays roughly constant up to Ni$_{47}$Al$_{53}$, see Fig.~\ref{f6}. 
If the toughness would be related to the hardness in a similar way than for Ni-rich NiAl, the fracture toughness of Ni$_{49}$Al$_{51}$ should be comparable to the toughness of composition between 45 and 41 at\% Al (using the literature data in Fig.~\ref{fxx3}) – which is not the case.  

\begin{figure}[]
	\center
	\includegraphics[width=0.5\textwidth]{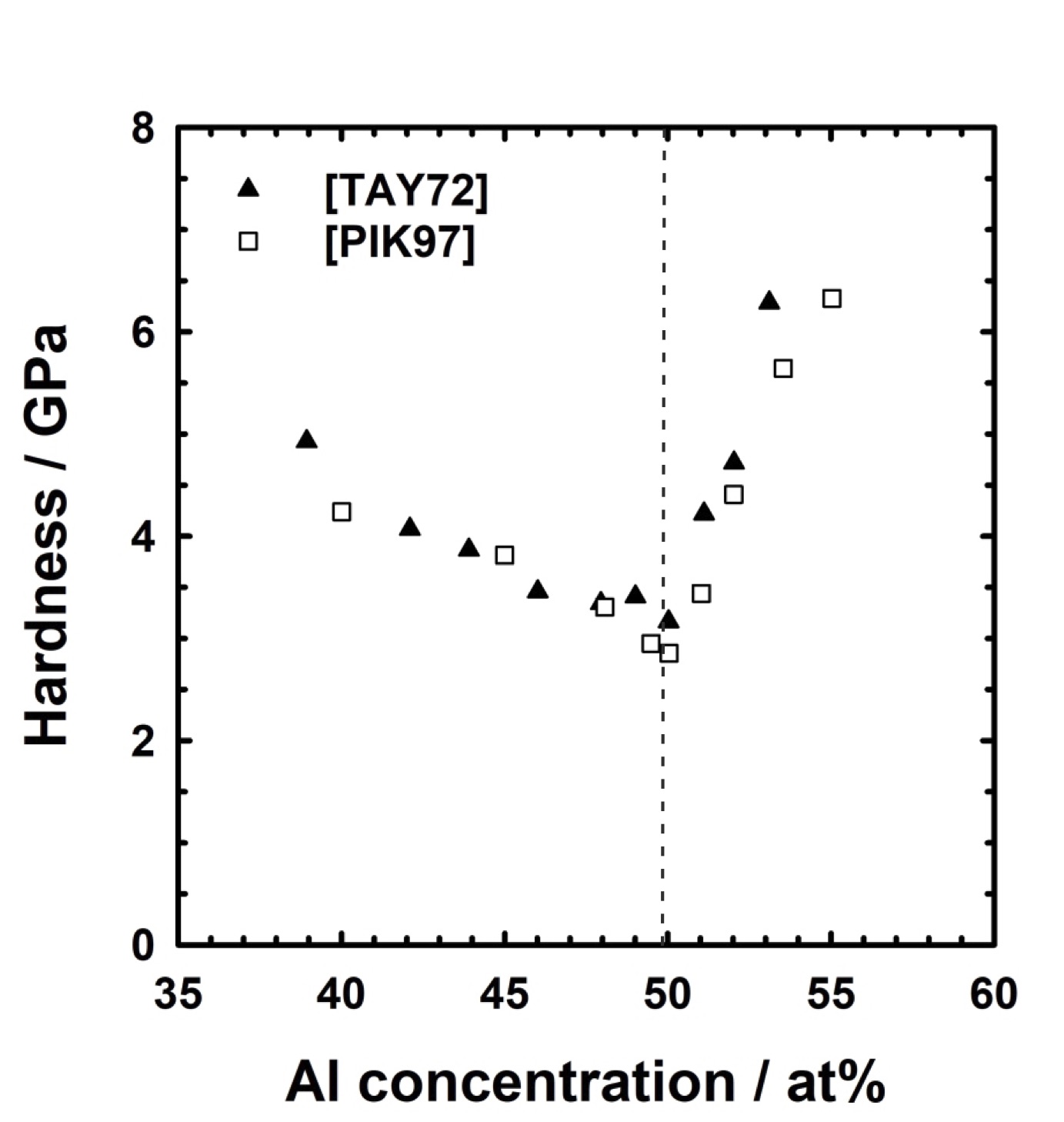} 
	\caption{\label{f3xx}
Hardness as a function of Al concentration of NiAl as determined by \cite{Pike1997,Taylor1972}.  
	}
\end{figure}

Clearly, an additional mechanism needs to be at play to explain the nearly concentration independent fracture toughness of Al-rich NiAl. 
The MD simulations, Fig.~\ref{f10}, show that the fundamental processes, namely crack propagation and dislocation emission are identical for the three studied compositions. 
In particular, there is no important difference in terms of the number or type of dislocations emitted and their distance travelled between the Ni$_{45}$Al$_{55}$ and the Ni$_{55}$Al$_{45}$ samples, and their Griffith strains  differ by just about 15\%.
%The strong hardening coefficient of structural vacancies is highlighted by the debris formed by the moving dislocations in Ni$_{45}$Al$_{55}$. 
In contrast to the other compositions, the Al-rich sample shows, however, a very low strain $\epsilon_c$ at which crack propagation starts, comparable to the theoretical strain $\epsilon_G$ according to Griffith’s energy balance. Crack propagation therefore seems to be favored by the presence of structural vacancies. 

As evidenced by Fig.~\ref{f14}, the crack in stoichiometric NiAl propagates by the nucleation and migration of kink pairs. 
Crack propagation by kinks was also observed in  \cite{Zhu2004c,Ker15}.
It can be therefore speculated that structural vacancies can act as sources for  kinks in the crack front, which then govern the dynamics of brittle crack propagation. 
This effect might be even more pronounced if the vacancies agglomerate to clusters as suggested by Gumbsch et al. \cite{Gumbsch1999}. 

Structural vacancies thereby not only impede dislocation motion but additionally facilitate crack propagation, leading to the observed nearly Al-concentration independent brittleness of Al-rich NiAl.        

\section{Summary}
The following conclusions on the composition dependent mechanical properties of non-stoichiometric B2 NiAl single crystals in the soft orientation can be drawn from our bending tests on notched microcantilevers and our atomistic fracture simulations:
\begin{itemize}
\item Off-stoichiometric NiAl shows a lower fracture toughness than stoichiometric NiAl, both in experiments and simulations, even though the theoretical fracture toughness according to the Griffith energy balance is comparable.
\item Plasticity is only observed in stoichiometric and Ni-rich NiAl, where the plastic contribution to fracture toughness decreases with increasing Ni-content. 
\item The fracture toughness of Al-rich NiAl is independent of the Al-concentration and with about 2 MPa m$^{1/2}$ lower than for the stoichiometric and Ni-rich compositions.
\item The relative values of the fracture toughness in Ni- and Al-rich NiAl and their different concentration dependence cannot be solely explained by the different hardening effect of Ni antisites (Ni-rich samples) and structural Ni vacancies (Al-rich samples), i.e. their influence on dislocation motion.
\item MD simulations show that crack propagation  in NiAl takes place by the nucleation and motion of kink pairs on the crack front. In addition to their role as potent obstacles for dislocations,
structural vacancies act as sources for crack front kink pairs. 
Thus rather than affecting the theoretical fracture toughness 
according to the thermodynamic Griffith criterion, structural vacancies in Al-rich NiAl lower the kinetic barrier for crack propagation.
\end{itemize}
In addition, the static, $K$-controlled simulations show that 
the usual approach for determining the fracture toughness, namely identifying the onset of crack propagation and the use of relatively short crack front segments, can not be applied to fracture in disordered systems. Here, new statistical approaches will be required.

% Here are two sample references: .
\section*{Acknowledgements}
PNB thanks C. Begau for his help in visualizing the dislocations in the atomistic simulations. 
The authors wish to acknowledge funding from the German Research Foundation (DFG) through the graduate schools1229 and 1896, the SFB/Transregio 103 (Single Crystal Superalloys) and the Cluster of Excellence ’Engineering of Advanced Materials’ at the Friedrich-Alexander-Universit\"at Erlangen-N\"urnberg (FAU) which is funded by the DFG within the framework of its ’Excellence Initiative’. 
PNB, SK and EB acknowledge funding from the European Research Council (ERC) under the European Union’s Horizon 2020 research and innovation programme (grant agreement No 725483).

\section*{Supplementary materials}
Supplementary material associated with this article can be found, in the online version, at XXX
\section*{References}

\bibliography{Mendeley_NiAl.bib}

\end{document}